\documentclass[twocolumn,aps,showpacs,preprintnumbers,nofootinbib, superscriptaddress, amsmath,amssymb]{revtex4}
\usepackage{mathtools}

\begin{document}

\author{I. I. Bondar}
\email{bondar.ivan@gmail.com}
\affiliation{Department of Physics, Uzhgorod National University, 54 Voloshyna St, Uzhgorod 88000, Ukraine}

\author{V. V. Suran}
\email{vasylsuran@gmail.com}
\affiliation{Department of Physics, Uzhgorod National University, 54 Voloshyna St, Uzhgorod 88000, Ukraine}

\author{D. I. Bondar}
\email{dbondar@princeton.edu}
\affiliation{Department of Chemistry, Princeton University, Princeton, NJ 08544, USA} 

\title{Multiphoton-double-ionization probability linearly depends on laser intensity: Experimental studies of barium}

\begin{abstract}
Despite inherently complex multiphoton dynamics, our observations show that Ba double ionization with an infrared laser (8800--8920cm$^{-1}$) resembles a single-photon process; namely, its probability is proportional to the laser intensity. In this regime, single-electron ionization is due to a four-photon resonant transition through the highly perturbed state $6p^2 \, {}^1D_2$, whereas double ionization is realized by the two-electron mechanism. Furthermore, we argue that these conclusions are valid for other alkaline-earth-metal atoms and other parameters of laser radiation. 
\end{abstract}

\date{\today} 

	
\maketitle 

\section{Introduction}
The effect of double ionization under the influence of laser radiation was originally observed in multiphoton ionization of Sr atoms with a Nd:glass laser \cite{Suran1975}. Henceforth, two-electron ionization of many other atoms has been observed \cite{Suran2009, Becker2012}, and diverse aspects of this phenomenon have been actively investigated ever since (see, e.g., Refs. \cite{Liontos2008, Motomura2009, Suran2009, Liontos2010, Liu2010}). In particular, it is well recognized that double ionization is ubiquitous among a broad variety of atomic groups \cite{Delone2000}. Moreover, the mechanism of two-electron ion generation significantly depends on the single-electron ionization regime as well as the spectral characteristics of the driving laser radiation.

When single-electron ionization occurs in the tunneling regime, doubly charged ions are obtained due to rescattering and collisional ionization by the liberated electrons. These electrons, produced by  ionization of neutral atoms in the first half-cycle of a laser pulse, return and collisionally ionize their parent ions after been accelerated in the second half-cycle \cite{Krausz2009, Becker2012}.

As far as ionization in the multiphoton regime is concerned, the number of photons needed for single-electron ionization determines the character of double ionization. The following two scenarios are possible: 

If only a few photons are required, then two-electron ions are generated by the \emph{sequential mechanism}: multiphoton ionization of singly charged ions preceded by ionization of the neutral atoms. This process takes place when atoms with low ionization potentials (e.g., alkaline-earth-metal atoms) are illuminated with laser radiation in the visible spectral range \cite{Delone2000}. 

The \emph{two-electron mechanism} is a non-sequential process of double ionization of alkaline-earth-metal atoms with infrared laser radiation. Note that rescattering, a widely studied phenomenon (see, e.g., Refs. \cite{Krausz2009, Becker2012}), cannot be a part of this process. Rescattering occurs once $r_e / r_a > 1$, where $r_a$ is the atomic radius and $r_e = F/\omega^2$ (the strength $F$ and frequency $\omega$ of the laser field in atomic units) is the oscillation amplitude of the free electron; however, $r_e / r_a \sim 0.1$ is typical for the two-electron mechanism.

To be ionized in this regime, an atom needs to absorb approximately twice as many photons as in the case of ionization with visible laser radiation. Abundant studies show that the two-electron mechanism of double ionization differs tremendously from the sequential one.  The main difference lies in the following: In the infrared spectral range, doubly charged ion generation is significantly higher than the anticipated yield for the sequential mechanism \cite{Bernat1991}. The resonant structure of the doubly charged ion yield reveals excitations of neutral atomic states, which are significantly perturbed by the ac Stark shift \cite{Suran2009}. However, in the case of ionization with visible laser radiation, the resonant structure of the doubly charged ion yield is due to excitations of singly charged ions \cite{Feldman1982, Agostini1984, Agostini1985, Dexter1985, Petite1986, Bondar1988, Nakhate1991, Tate1991, Haugen1992, Suran2005, Suran2009}. Furthermore, studies involving auxiliary ionization and excitation \cite{Suran2009} demonstrate that infrared double ionization feeds on neutral atoms, but not on single-electron ions. All these facts indicate that infrared ionization of alkaline earth atoms is governed by the two-electron mechanism of double ionization, when the ions are generated directly from the neutral atoms. It is assumed that doubly charged ions are obtained as a result of the detachment of the two outermost electrons from atoms excited to autoionizing states, whose energies approach the second ionization potential \cite{Bondar2003}. These states are populated by one-photon jumps through low-lying autoionizing states. 

In the current work, we present experimental results for the dependence of the probability of doubly charged ion formation on the laser field intensity for atomic Ba ionization with an infrared color-center laser. We use a common set-up employed widely in multiphoton studies and described in detail elsewhere (see, e.g., Refs. \cite{Delone2000, Ammosov1992}). Briefly, our set-up consists of a laser beam directed into a vacuum chamber, where it intersects a Ba atomic beam. Production of Ba$^{+}$ and Ba$^{2+}$ ions occurs at the intersection of the beams. Ion signals are then measured and analyzed by mass spectrometry. The concentration of neutral Ba atoms in the beam is about $10^{11}$ cm$^{-3}$. We utilize a linearly polarized color-center laser with the frequency ($\omega$) in the region 8800--8920cm$^{-1}$ and the pulse duration $\tau \approx 4 \times 10^{-8} s$. The line width of laser generation is 4cm$^{-1}$. Under these  conditions, single-electron ionization of Ba requires five-photon absorption. We experimentally measure the Ba$^{+}$ and Ba$^{2+}$ yields by varying the laser radiation frequency (Sec. \ref{Sec_freq_dep}) and intensity (Sec. \ref{Sec_intesity_dep}). Analyzing measured data (Sec. \ref{Sec_double_ion_prob}), we conclude that the probability of Ba double ionization linearly depends on the laser intensity. We then formulate the rule of thumb that extends our conclusions to other cases (Sec. \ref{Sec_conclusion}).

\section{LASER FREQUENCY DEPENDENCE OF Ba$^+$ AND Ba$^{2+}$ YIELDS}\label{Sec_freq_dep}

\begin{figure}
	\begin{center}
		\includegraphics[scale=0.6]{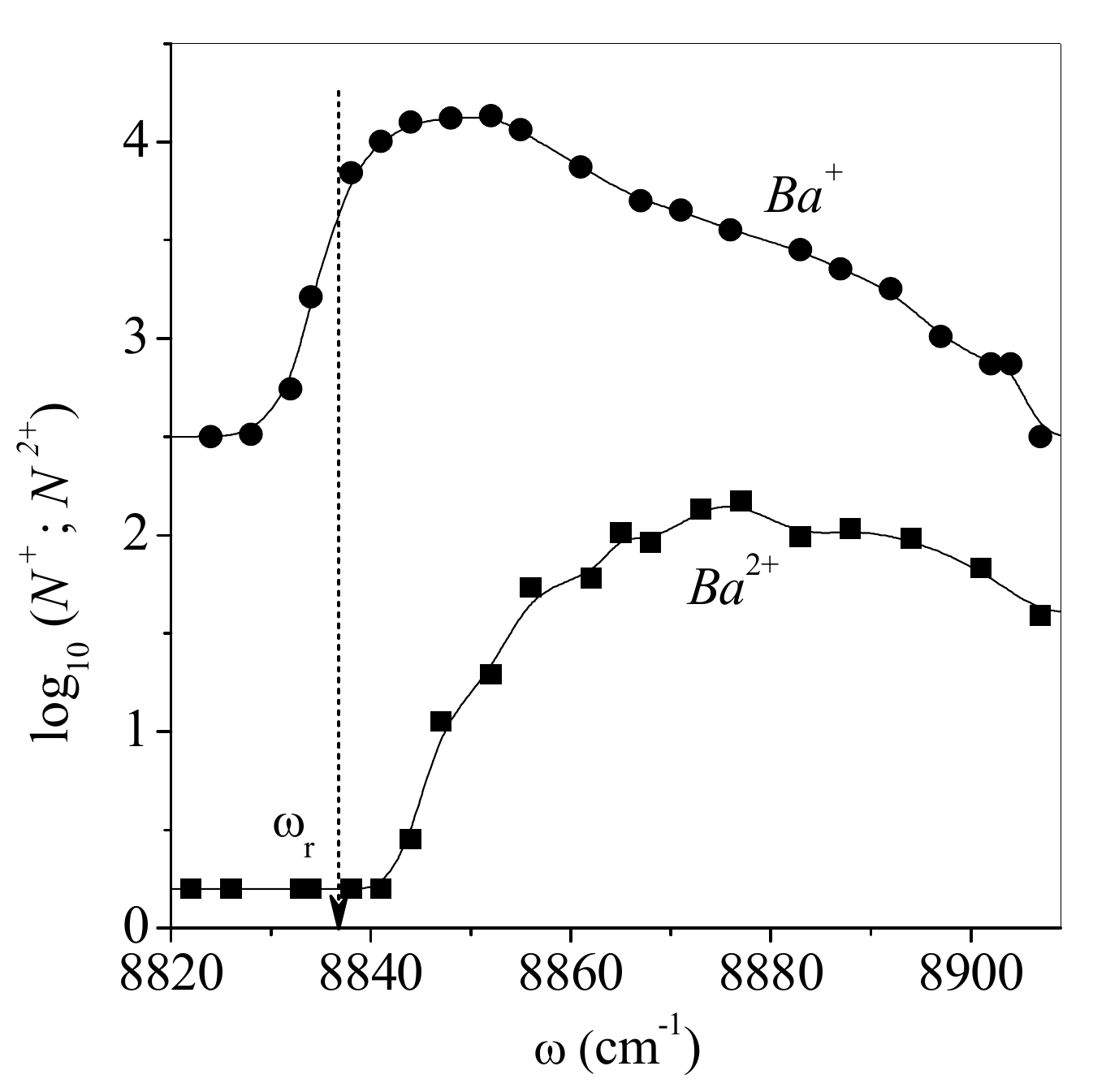}
		\caption{Dependences of the singly and doubly charged ion yields on the laser radiation frequency for Ba atom ionization with color-centre laser radiation. The dashed  vertical line denotes the frequency of the four-photon excitation of the unperturbed state $6p^2 \, {}^1D_2$ ($\omega_r = 8836$cm$^{-1}$).}
		\label{Fig1}
	\end{center}
\end{figure}

Figure \ref{Fig1} presents the results of studies of the laser frequency dependencies of Ba$^+$ and Ba$^{2+}$ generation. The laser intensity was maintained constant with the laser power density in the interaction region $P \approx 5 \times 10^{10}$W/cm$^2$. Figure \ref{Fig1} reveals a resonant structure in Ba$^+$ and Ba$^{2+}$ yields. This result qualitatively coincides with our previous results \cite{Bondar1998, Bondar2000} in the same spectral region but with different intensities. An analysis shows that the resonant maximum in Ba$^+$ yield, displayed in Figure \ref{Fig1}, is due to five-photon ionization of Ba atoms through the four-photon resonance with state $6p^2 \, {}^1D_2$ substantially perturbed by the ac Stark effect. This interpretation is confirmed by the resonance's asymmetry, its large width, and the observation that its peak is detuned from the resonant frequency $\omega_r = 8836$cm$^{-1}$ of four-photon excitation of the unperturbed state $6p^2 \, {}^1D_2$.  In our experiment, the energy shift of the $6p^2 \, {}^1D_2$ state is $\Delta E \approx 250cm^{-1}$; this value is estimated from the resonance's asymmetry and width.

\begin{figure}
	\begin{center}
		\includegraphics[scale=0.7]{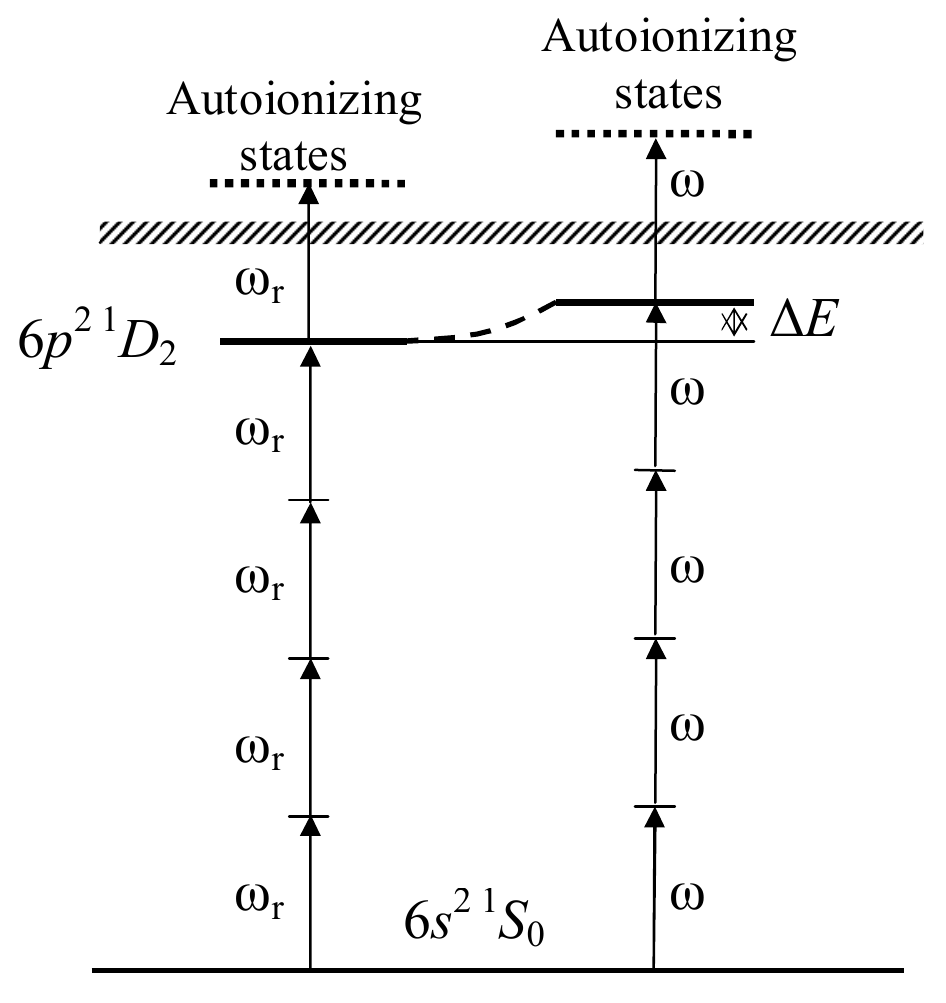}
		\caption{Scheme of the perturbed-state $6p^2 \, {}^1D_2$ excitation.}
		\label{Fig2}
	\end{center}
\end{figure}

The scheme of the dynamical resonance realization involving the perturbed state $6p^2 \, {}^1D_2$ is presented in figure \ref{Fig2}. The shape and position of the resonant maximum, perturbed by the ac Stark effect, depend on the following quantities: The dynamical polarizabilities  for the ground $\alpha_0$ and resonant $\alpha_n$ states, the resonant ionization probability, and the spatial-temporal distribution of the laser radiation in the interaction region. In our case, the laser radiation frequencies differ significantly from the frequencies of single-photon transitions from the ground $6s^2 {}^1S_0$ and excited $6p^2 \, {}^1D_2$ states to any other state of the Ba atom; additionally, the frequency variation of the laser radiation is much narrower than intra-resonance distances among such transitions.  Hence, the dynamical polarizabilities of the ground and excited states should not strongly depend on the frequency of the utilized laser radiation. Also, we have  previously measured the dynamical polarizability of the $6p^2 \, {}^1D_2$ state in the spectral range around $\omega \sim 8850$cm$^{-1}$ \cite{Bondar1998c}; the obtained value is $\alpha_n \approx -1.8\times 10^3$ a.u. Since, the laser radiation frequencies are significantly lower than the frequency of the single-photon transition from the ground to the first resonant state ($\omega = 18060$cm$^{-1}$), the value of the ground stateÕs dynamical polarizability should not differ significantly from the corresponding DC polarizability ($\alpha_0 \approx 280$ a.u.). Therefore, the shape of the resonance in Ba$^+$ yield is manly due to the perturbed $6p^2 \, {}^1D_2$ state excitation.

\begin{figure}
	\begin{center}
		\includegraphics[scale=0.6]{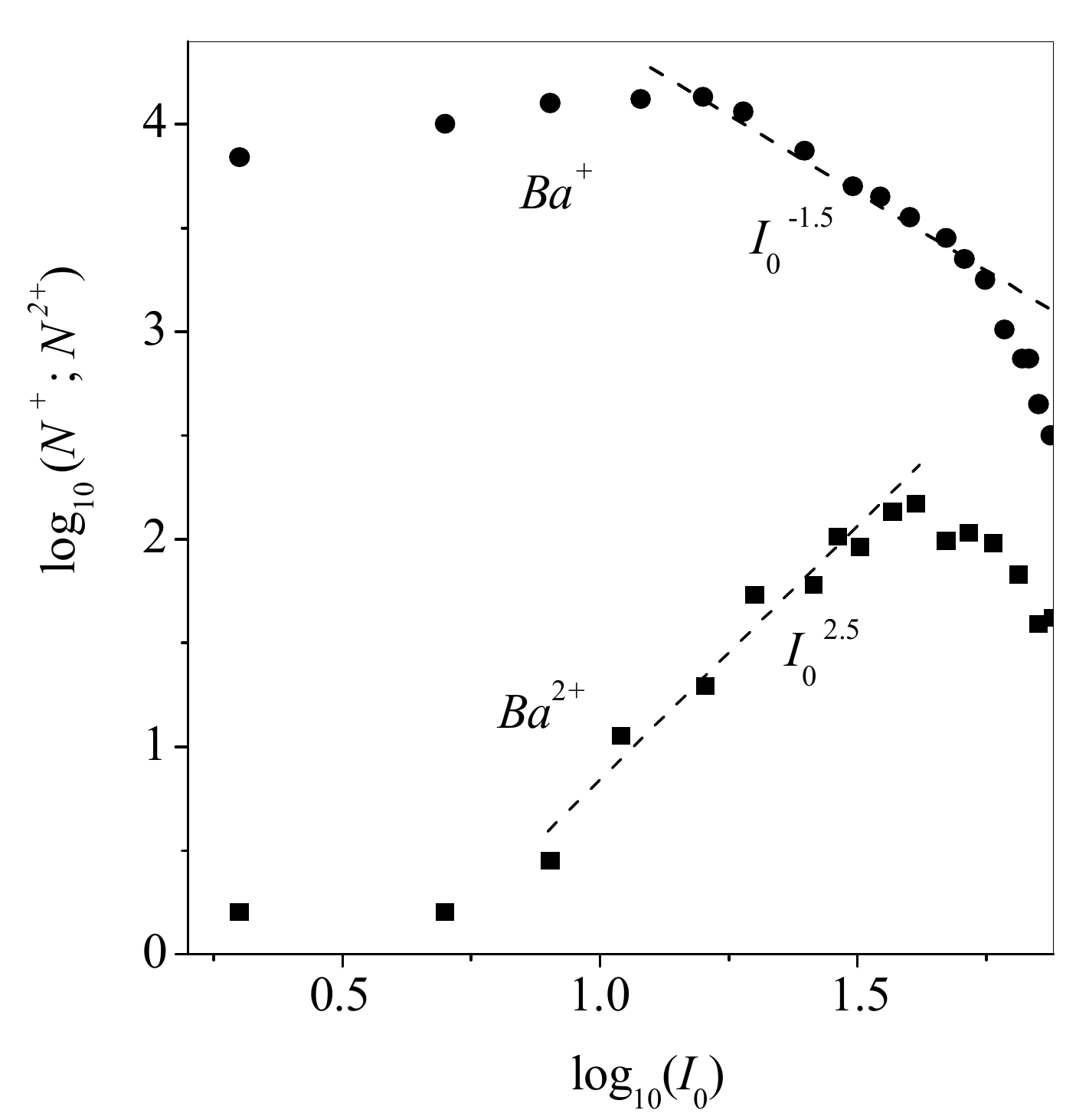}
		\caption{Dependence of Ba$^+$ and Ba$^{2+}$ yields on the laser radiation frequency re-plotted in the coordinates ($N^+$, $N^{2+}$; $I_0$).}
		\label{Fig3}
	\end{center}
\end{figure}

As shown above, the dynamical polarizabilities $\alpha_0$ and $\alpha_n$ are nearly frequency independent in our experiment. In this case, utilizing the connection between the frequency $\omega$ and the intensity $I_0$ :  $I_0 \sim \omega-\omega_r$, the dependence of the Ba$^+$ yield on $\omega$ can be translated to the yield dependence on the laser intensity $I_0$, for which the dynamical resonance occurs. The variation range of the intensity $I_0$, corresponding to the resonant maximum in Ba$^+$ yield, is bounded by the peak intensity of a laser pulse, $I_0 = I$. The Ba$^+$ and Ba$^{2+}$ yield dependencies on the laser frequency, obtained in this way, are depicted in figure \ref{Fig3}. 

In a single laser pulse, the process of ionization of Ba atoms through the dynamical resonance with the excited state $6p^2 \, {}^1D_2$ becomes efficient when the laser intensity is close or equal to $I_0$. Then, Ba$^+$ formation is a threshold process with respect to the laser intensity. Before dynamical resonance is achieved, Ba ionization is mainly realized by a direct five-photon process. Utilizing typical values for the effective cross sections of direct five-photon ionization \cite{Ammosov1992}, qualitative estimates show that direct ionization of Ba atoms, before the dynamical resonance, is far from saturation in our experiments. Hence, the concentration of neutral Ba atoms when the laser intensity reaches $I_0$ does not differ significantly from the initial concentration ($n_0$).

The duration of Ba$^+$ production through the dynamical resonance with the perturbed state $6p^2 \, {}^1D_2$ coincides with the tuning time $\Delta \tau$ of the dynamical resonance. Since $\Delta \tau$ is significantly shorter than the laser pulse duration $\tau$, the laser field intensity does not differ much from $I_0$ within the time interval $\Delta \tau$, i.e., Ba$^+$ ions are generated by a laser field of the constant intensity $I_0$.

To analyze the obtained results, we assume that within the interaction volume of the laser field with the atomic beam, the temporal--spatial distribution of the laser field intensity is Gaussian \cite{Delone2000}
\begin{align}\label{eq1}
	I_L (r,z,t) = \frac{I}{\left[ 1 + (z/z_0)^2 \right]^{2}} \exp\left\{  \frac{-2(r/r_0)^2}{ 1 + (z/z_0)^2} -2 \left(\frac{t}{\tau}\right)^2 \right\},
\end{align}
where $z$ and $r$ are the spatial coordinates, $r_0$ is the minimal radius in the focus; $z_0 = \pi r_0^2 / \lambda$ is the Rayleigh length, and $\lambda$ is the radiation wave length.

\begin{figure}
	\begin{center}
		\includegraphics[scale=0.6]{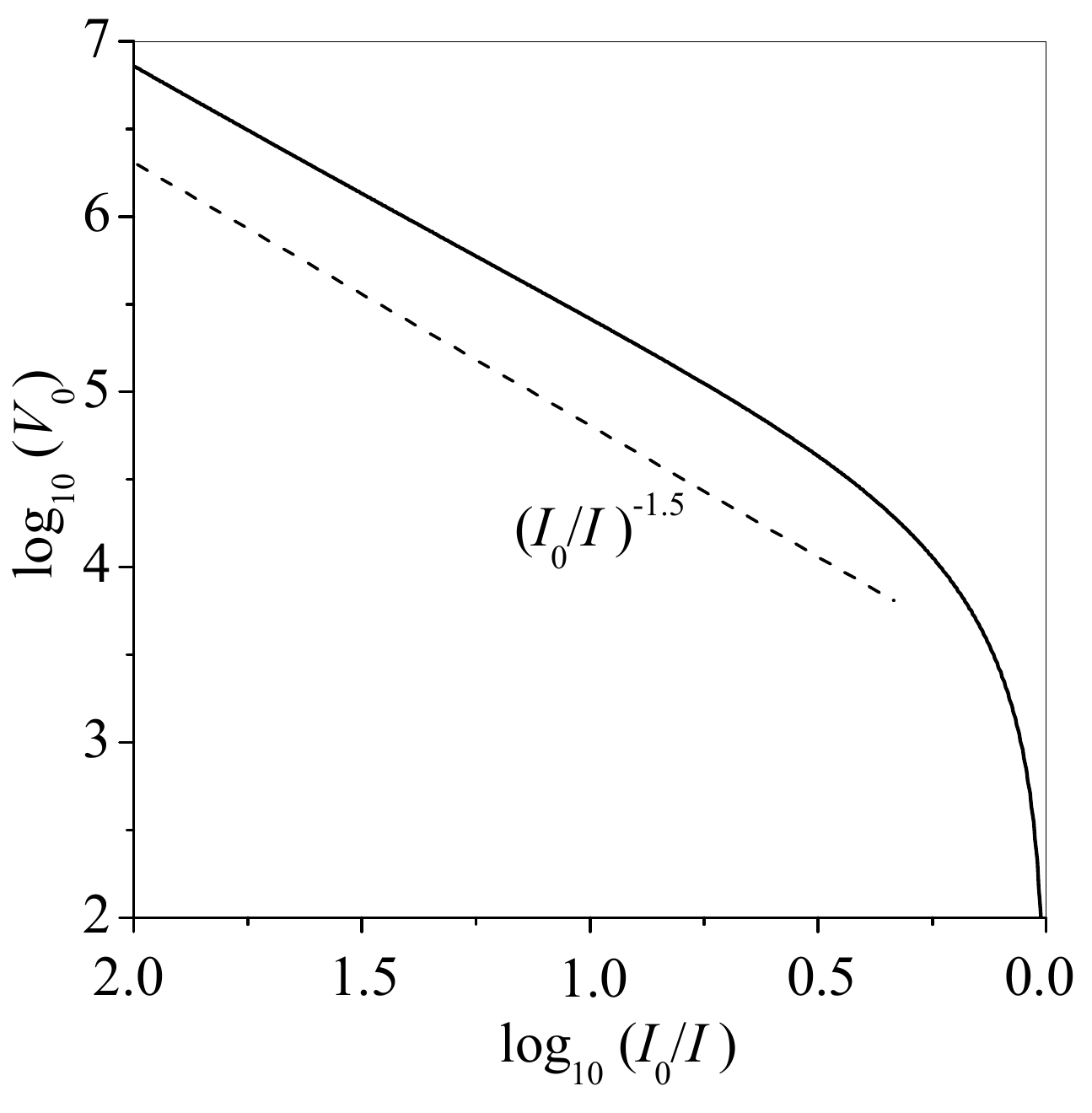}
		\caption{The function $V_0(I_0)$ given by equation (\ref{eq3}).}
		\label{Fig4}
	\end{center}
\end{figure}

In our experiments, Ba$^+$ ions are produced at all those points of the interaction volume where the laser intensity reaches $I_0$ within a laser pulse duration. The region of these points is bounded by a surface, where the intensity $I_0$ is maximal ($I_0=I$). According to equation (\ref{eq1}), this surface is given by
\begin{align}\label{eq2}
	r(z) = r_0 \sqrt{  \frac{1 + (z/z_0)^2}{2} \ln \frac{I/I_0}{1 + (z/z_0)^2}  },
	\, z_1 \leq z \leq z_2,
\end{align}
where $z_{1,2} = \pm z_0 \sqrt{ I/I_0 - 1}$. The volume of this region ($V_0$) is \cite{Bondar2004}
\begin{align}\label{eq3}
	V_0\left( I/I_0 \right) =& \frac{2\pi^2 r_0^4}{\lambda} \left[ \frac{1}{9} \left( I/I_0 - 1 \right)^{3/2} \right. \notag\\
	& \left. +\frac{2}{3} \sqrt{I/I_0 - 1} - \frac{2}{3} \arctan\sqrt{I/I_0 - 1}  \right]. 
\end{align}
Note that for the experimental results in figure \ref{Fig3}, the intensity $I_0$ varies whereas $I$ is constant. The function $V_0(I_0)$ is depicted in figure \ref{Fig4}.

When the laser frequency is detuned from the resonant frequency $\omega_r$, the dynamical resonance tuning occurs at a higher laser intensity $I_0$. Hence, the excitation probability of the perturbed state $6p^2 \, {}^1D_2$ increases, while the interaction volume $V_0$, where the laser intensity reaches $I_0$, shrinks. As a result, the Ba$^+$ yield enhancement in figures \ref{Fig1} and \ref{Fig3}, when the radiation frequency is offset from $\omega_r$, is due to the increase in the excitation probability of the state $6p^2 \, {}^1D_2$, whereas the subsequent ion yield suppression is caused by the contraction of the volume $V_0$.

The analysis of our experiment shows that ionization of Ba atoms through the resonance with the perturbed state $6p^2 \, {}^1D_2$ is saturated under the condition of a significant detuning from the resonance frequency $\omega_r$ (for a large value of $I_0$). If the ionization process is saturated, then the dependence of the Ba$^+$ yield on the intensity $I_0$ should be proportional to the dependence of the volume $V_0$ on the intensity. According to equation (\ref{eq3}) and figure \ref{Fig4}, the function $V_0(I_0)$ has an asymptotic $V_0 \sim I_0^{-1.5}$ for $I_0 \ll I$. Therefore, under the condition of Ba ionization saturation through the resonance with the state $6p^2 \, {}^1D_2$ when the resonance tuning occurs for laser intensities $I_0 \ll I$, the Ba$^+$ ion yield should be $N^+ \sim I_0^{-1.5}$.

A plot of the power law $N^+ \sim I_0^{-1.5}$ is shown in figure \ref{Fig3} for comparison. It shows that after the maximum yield is achieved, the dependence of the Ba$^+$ yield on the intensity $I_0$ is indeed approximated by $N^+ \sim I_0^{-1.5}$. Thus, Ba ionization through the resonance with the perturbed state $6p^2 \, {}^1D_2$ occurs in the saturation regime for the intensities $I_0$ corresponding to a decline in the ionization yield. Additionally, the successful approximation of the ion yield by the power law $N^+ \sim I_0^{-1.5}$ justifies the employment of the Gaussian distribution for the laser intensity to interpret the experimental results.

Note that a saturation of Ba$^+$ production is also observed for large values of $I_0$. A deviation of  $N^+(I_0)$ from the power law $N^+ \sim I_0^{-1.5}$ for large $I_0$ can be attributed to a significant decrease of the volume $V_0(I_0)$ in the neighborhood of the intensity $I_0= I$ (see figure \ref{Fig4}).

Consider now the results for the Ba$^{2+}$ yield. According to figure \ref{Fig3}, these ions are effectively formed when Ba$^+$ generation through the resonance with the perturbed state $6p^2 \, {}^1D_2$ is saturated. The Ba$^{2+}$ yield first considerably increases and then decreases for the monotonically increasing intensity $I_0$. Moreover, the Ba$^{2+}$ yield declines for those intensities $I_0$ where Ba$^+$ generation falls more rapidly than the dependency $N^+ \sim I_0^{-1.5}$, due to the  above-mentioned volume $V_0$ shrinkage in the vicinity $I_0 = I$.

The observation that the rapid yield declines for Ba$^+$ and Ba$^{2+}$ took place in the same intensity $I_0$ interval indicates that the processes of Ba$^{2+}$ and Ba$^+$ generation not only are connected with the dynamical resonant with the state $6p^2 \, {}^1D_2$, but also occur in the same interaction region, whose volume is given by equation (\ref{eq3}).

According to figure \ref{Fig3}, the Ba$^{2+}$ yield enhancement for increasing intensity $I_0$ can be approximated by $N^{2+} \sim I_0^{2.5}$.  As shown above, the volume $V_0(I_0)$ decrease with increasing $I_0$ and for intensities enhancing the Ba$^{2+}$ yield, $V_0 \sim I_0^{1.5}$. Taking into account these dependencies, we obtain the probability of Ba$^{2+}$ generation as a function of the laser intensity: $W^{2+} = N^{2+} / V_0 \sim I_0^4$. Excitation of the state $6p^2 \, {}^1D_2$  is a four-photon process, and thus, its probability is characterized by the same power law. This fact additionally indicates that Ba double ionization is closely connected with the dynamical resonance with the perturbed state $6p^2 \, {}^1D_2$. 

We remind the reader that in all precedding discussion, the intensity ($I_0$) of the dynamical resonance tuning is varied, while the peak laser intensity ($I$) remains fixed.

\section{LASER INTENSITY DEPENDENCE OF Ba$^+$AND Ba$^{2+}$ YIELDS}\label{Sec_intesity_dep}

\begin{figure}
	\begin{center}
		\includegraphics[scale=0.6]{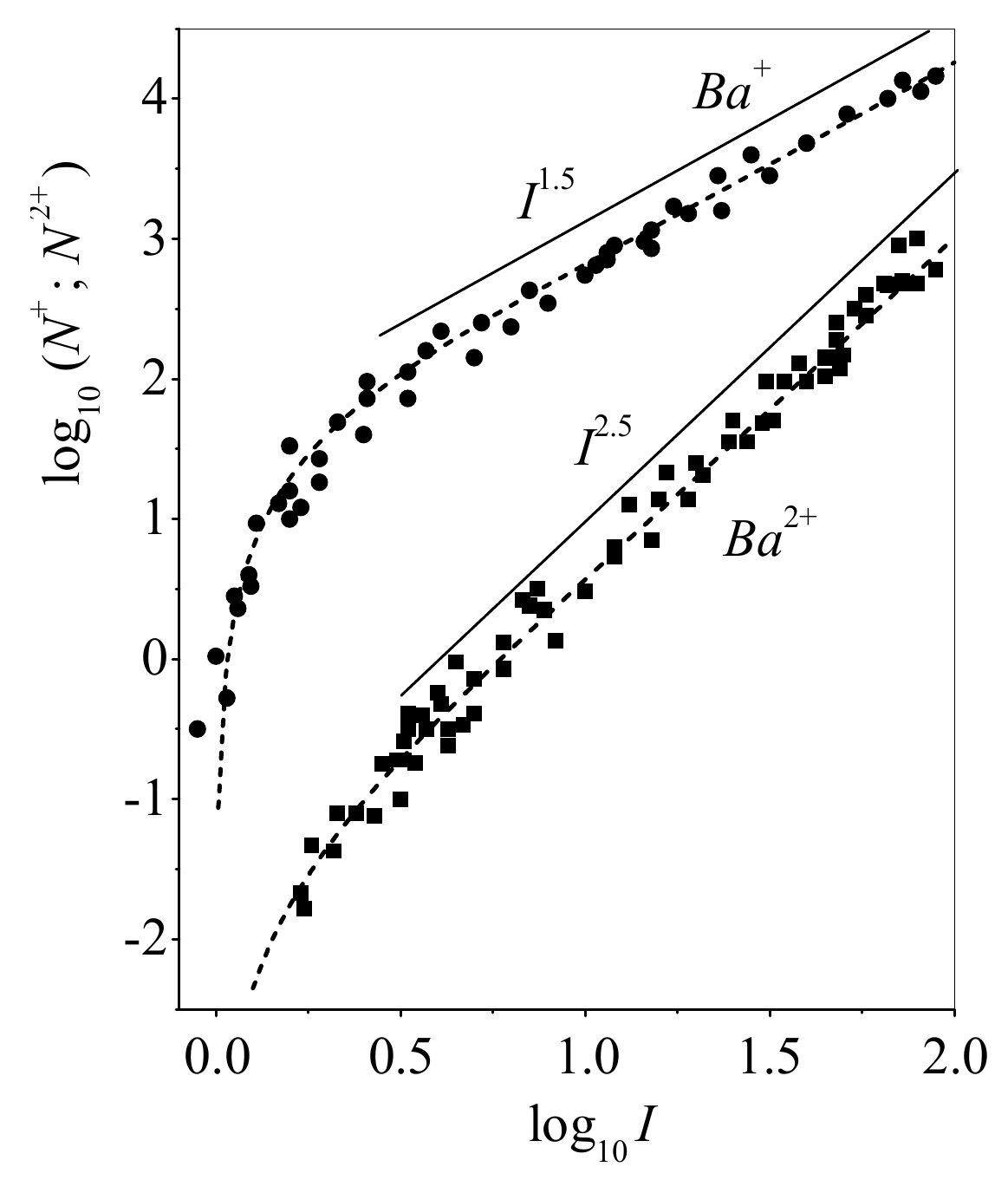}
		\caption{Dependence of the Ba$^+$ and Ba$^{2+}$ yields on the intensity $I$ of color-centre laser radiation at the frequency $\omega=8865$cm$^{-1}$. Points denote experimental measurements. Dashed lines represent experimental data fits by $V_0(I)$ [equation (\ref{eq3})] for Ba$^+$ and by $IV_0(I)$ for Ba$^{2+}$, respectively. For comparison, the power laws $N^+ \sim I^{1.5}$ and $N^{2+} \sim I^{2.5}$  are displayed by solid lines.}
		\label{Fig5}
	\end{center}
\end{figure}

We have also studied Ba$^+$ and Ba$^{2+}$ formation by varying the laser intensity $I$ while maintaining the fixed laser frequency $\omega=8865$cm$^{-1}$. This frequency belongs to the regions in figure \ref{Fig3} where the functions $N^+(I_0)$ and $N^{2+}(I_0)$ can be approximated by the power laws $N^+ \sim I_0^{-1.5}$ and $N^{2+} \sim I_0^{2.5}$, respectively. Obtained results are depicted in figure \ref{Fig5}.

\begin{figure}
	\begin{center}
		\includegraphics[scale=0.6]{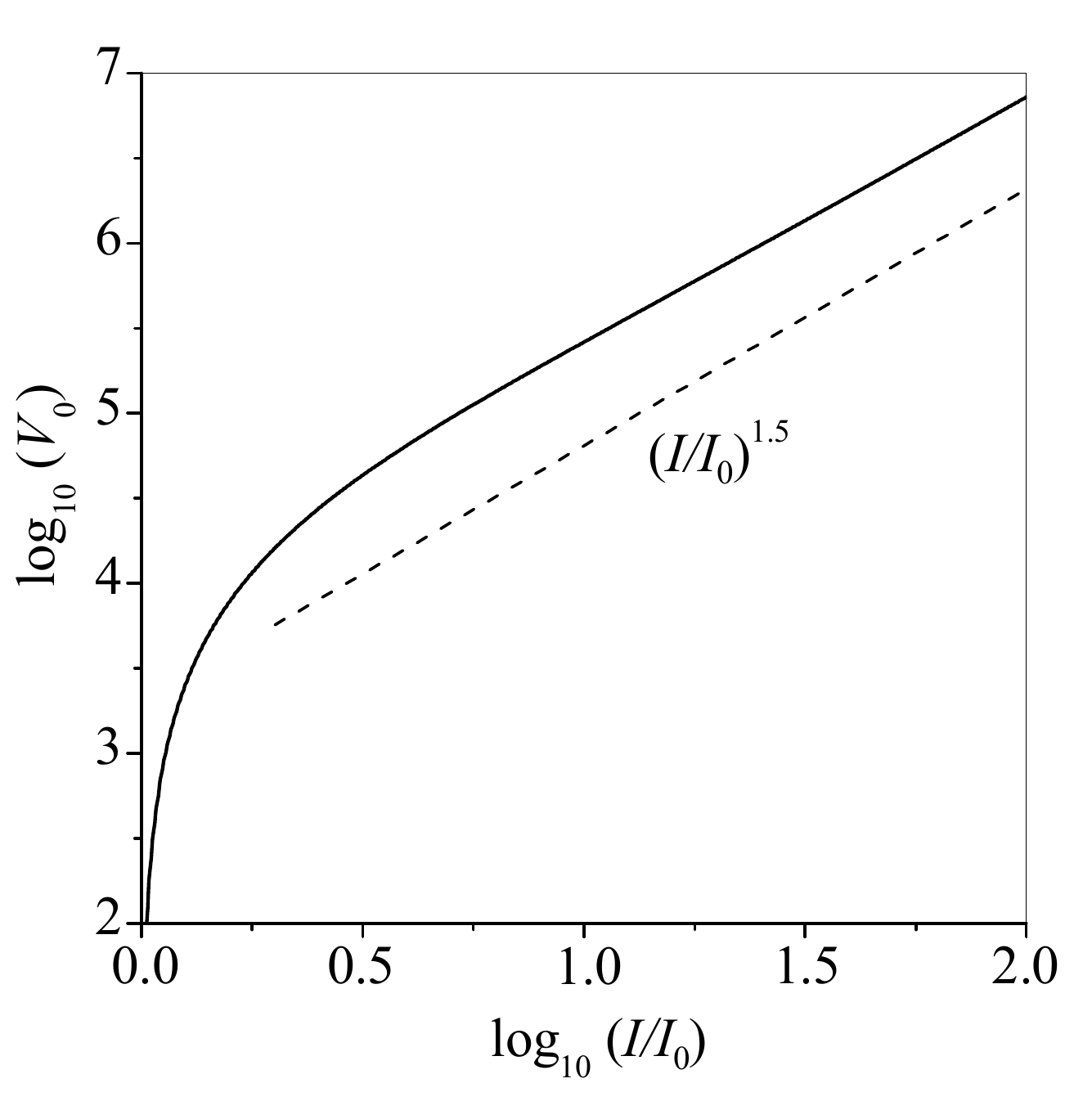}
		\caption{The function $V_0(I)$ given by equation (\ref{eq3}).}
		\label{Fig6}
	\end{center}
\end{figure}

Consider first the Ba$^+$ yield as a function of the intensity $I$. According to figure \ref{Fig5}, such a dependence can be fitted for a wide range of laser intensities by the power law $N^+ \sim I^{1.5}$, indicating that Ba$^+$ generation is saturated. Since Ba ionization in this case occurs through the resonance with the perturbed state $6p^2 \, {}^1D_2$, the Ba$^+$ yield should be proportional to the volume $V_0$ where the intensity $I_0$ is reached. The volume $V_0$ as a function of $I$ for a Gaussian laser beam is also given by equation (\ref{eq3}). Since measurements of $N^+(I)$  were performed at a fixed laser frequency, the intensity $I$ in equation (\ref{eq3}) is a variable whereas $I_0$ is a constant. The plot of the function $V_0(I)$ is shown in figure \ref{Fig6}. According to equation (\ref{eq3}) as well as figure \ref{Fig6}, the dependence of $V_0$ on $I$ can be approximated by $V_0 \sim I^{1.5}$. Therefore, the experimentally measured function $N^+(I)$ is proportional to $V_0(I)$, manifesting Ba ionization saturation.

When we begin increasing the laser intensity $I$, the dynamical resonance tuning initially occurs at the middle of a laser pulse. A further intensity increase shifts the resonance tuning time ($t_0$) to the beginning and end of a laser pulse. 

It follows from equation (\ref{eq1}) that the time $t_0$, when the laser intensity at a point $(r,z)$ reaches $I_0$, equals
\begin{align}\label{eq4}
	t_0(r,z) = \pm \sqrt{ \frac{1}{2}\ln\frac{I/I_0}{1 + z/z_0} - \frac{ (r/r_0)^2 }{1 + (z/z_0)^2} }. 
\end{align}
Utilizing equation (\ref{eq4}), we calculate the time ($T$) averaged over the volume where the laser intensity rises to $I_0$ (i.e., the resonance tuning time) for a Gaussian laser beam:
\begin{align}\label{eq5}
	T (I/I_0) &= \frac{4\pi}{V_0(I/I_0)} \int_{z_1}^{z_2} dz \int_0^{r(z)} r dr \left|t_0(r,z)\right| \notag\\
		&= \frac{8\pi^2 r_0^4 \tau}{3\lambda V_0(I/I_0)} \sqrt{ I/I_0 -1 } q( I/I_0 -1 ),
\end{align}
where
\begin{align}
	q(x) = \int_0^1 dy \, \left(1 + xy^2 \right) \left( \frac{1}{2} \ln \frac{1+x}{1+xy^2} \right)^{3/2}.
\end{align}
Note that the time $T$ in equation (\ref{eq5}) is measured from the laser pulse center.

\begin{figure}
	\begin{center}
		\includegraphics[scale=0.6]{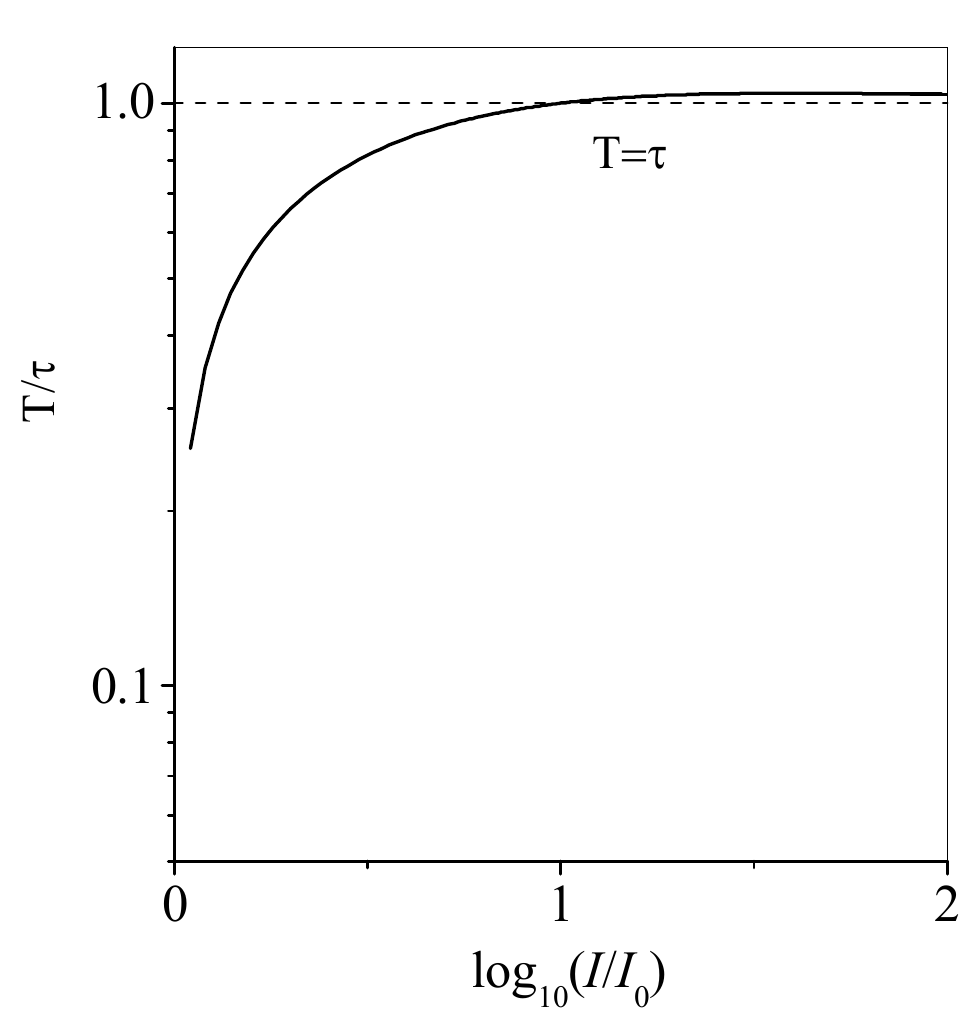}
		\caption{Dependence of the time $T$ averaged over volume [equation (\ref{eq5})], when the instantaneous laser intensity reaches $I_0$, on the laser radiation intensity $I$.}
		\label{Fig7}
	\end{center}
\end{figure}

A plot of the function $T(I)$ is shown in figure \ref{Fig7}, which illustrates that the averaged time $T$ is monotonically increasing and  approaches a constant value $\approx \tau$ for large $I$. Moreover, the equality $T(I) \approx \tau$ occurs for those laser intensities where the dependence $V_0(I)$ is well approximated by the power law $V_0 \sim I^{1.5}$. Hence, the tuning of the dynamical resonance with the state $6p^2 \, {}^1D_2$ occurs on average at the beginning of a laser pulse.

As was shown above, the duration of Ba$^+$ formation is approximately equal to the duration of the resonant tuning and is significantly shorter than the laser pulse duration. Taking this fact into account, we conclude that Ba$^+$ ion generation for the intensities $I \gg I_0$ takes place on average at the beginning of the laser pulse.

Ba atom excitation to low-lying autoionizing states occurs simultaneously with Ba$^+$ formation. A high density of the autoionizing states in the energy neighborhood of five-photons (see, e.g., Refs. \cite{Kozlov1981, Kotochigova1987, Bente1989}) suggests the possibility of Ba atom excitations to these states. The corresponding excitation probability should be rather high because this excitation is realized through the intermediate dynamical resonance with the bound state $6p^2 \, {}^1D_2$.

Since Ba$^+$ generation is saturated in our experiments, the ionization duration and the duration of Ba excitation to the autoionizing states coincide and are approximately equal to the duration of the resonant tuning with the state $6p^2 \, {}^1D_2$. In other words, both ionization and excitation occurs at a nearly constant laser intensity $\approx I_0$ at all points in the interaction volume. As a result, the concentrations of Ba$^+$ ions and excited Ba atoms in autoionizing states are uniform throughout the interaction volume and do not depend on the laser intensity $I$.

Examine now the dependence of Ba$^{2+}$ yield on $I$. According to figure \ref{Fig5}, increasing the intensity enhances the Ba$^{2+}$ yield, at first rapidly and then more slowly, such that the power law $N^{2+} \sim I^{2.5}$ is valid for large $I$. Note that this approximation is valid for those intensities for which the singly charged ion yield is given by $N^+ \sim I^{1.5}$.

\section{DEPENDANCE of Ba$^{2+}$ FORMATION PROBABILITY ON LASER INTENSITY}\label{Sec_double_ion_prob}

The rate of Ba$^{2+}$ generation (i.e., the probability per unit time) as a function of the intensity can be determined from the measured function $N^{2+}(I)$ displayed in figure \ref{Fig5}. To simplify the analysis, we consider the laser intensity interval in which the approximation $N^{2+} \sim I^{2.5}$ is valid. The most general expression for the probability of double ionization per unit time reads
\begin{align}\label{eq6}
	w^{2+} = N^{2+} / (nV\Delta t),
\end{align}
where $N^{2+}$ is the number of ions, $n$ is the concentration of target particles for ion formation, $V$ is the volume of ion production, and $\Delta t$ is the duration of the process. 

Ba$^+$ ions and Ba atoms excited to autoionizing states fuel  the sequential and two-electron  double ionization mechanisms, respectively.  Ba$^{2+}$ creation takes place wherever singly charged ions or neutral atoms are present. As shown above, Ba$^+$ ions and Ba atoms in autoionizing states are produced where the laser intensity reaches the value $I_0$ within a pulse duration. In this case, the sequential and two-electron mechanisms are spatially indistinguishable. The volume $V_0$ of such points is a function of the radiation intensity $I$ given by equation (\ref{eq3}). The power law $V_0 \sim I^{1.5}$ approximates $V_0(I)$ for large $I$. Hence, the volume $V$ in equation (\ref{eq6}) must be equal to $V_0$ and $V \sim I^{1.5}$.

As discussed above, in our experiments (see figure \ref{Fig5}) the concentrations of Ba$^+$ ions and Ba atoms in the autoionizing states do not depend on the intensity and are constant in the whole volume where Ba atom is ionized through the resonant with the perturbed $6p^2 \, {}^1D_2$ state. Namely, $n=const$ in equation (\ref{eq6}).

Moreover, both Ba$^+$ ions and excited Ba atoms are formed on average at the beginning of a laser pulse, and subsequently have the whole laser pulse to interact with. Therefore, the averaged time for Ba$^{2+}$ production equals the laser pulse duration and does not depend on $I$, i.e., $\Delta t= const$ in equation (\ref{eq6}).

Finally, having substituted all these dependences ($N^{2+} \sim I^{2.5}$, $V \sim I^{1.5}$, $n=const$, $\Delta t = const$) into equation (\ref{eq6}), we obtain that the double ionization rate linearly depends on the laser intensity ($w^{2+} \sim I$) for large $I$ ($I \gg I_0$). This result cannot be explained by the sequential mechanism of Ba$^{2+}$ formation since straightforward multiphoton ionization of Ba$^+$ cannot be linearly dependent on the laser intensity. Indeed, experimental studies \cite{Suran2009} on double ionization of alkaline-earth-metal atoms by visible laser radiation show that the doubly charged ion yield of the sequential mechanism is $\sim I^N$, where $N$ is the minimum number of quanta needed to ionize singly charged ions.

Contrary to the sequential dynamics, the two-electron mechanism of Ba$^{2+}$ formation, which is due to one-photon jumps through the spectrum of autoionizing states, agrees well with the observed linear dependence on the intensity $I$. The autoionizing states are populated and transitions between them are induced when the laser intensity reaches the value of $I_0$. The widths of the autoionizing states formed by the one-photon transitions may be broader than the decay widths of these states. Such conditions ensure a high probability of doubly charged ion production. 

\section{Conclusions}\label{Sec_conclusion}

The presented experimental results suggest that the two-electron mechanism is responsible for double ionization of Ba atoms by infrared laser radiation. The Ba double ionization probability is shown to depend linearly on the laser radiation intensity. This counterintuitive dependency is attributed to the the saturation of the four-photon transition from the ground state to $6p^2 \, {}^1D_2$ and the non-saturation of single-photon transitions between autoionizing states.

Furthermore, these conclusions are valid for other alkaline earth atoms as well as other parameters of laser radiation. The current and previous studies on Sr and Ba in infrared laser radiation (see, e.g., Ref. \cite{Suran2009}) can be summarized as the following \emph{rule of thumb}: Whenever the dependence of the singly charged ion yield $N^+$ on the laser radiation intensity $I$ is $N^+ \sim I^{1.5}$, the doubly charged ion yield of the two-electron mechanism is $N^{2+} \sim I^{2.5}$. Furthermore, the analysis employed in the current work is applicable to such cases and leads to the conclusion that the probability of double ionization is proportional to the laser intensity.

\bibliography{Bondar}

\begin{thebibliography}{29}
\expandafter\ifx\csname natexlab\endcsname\relax\def\natexlab#1{#1}\fi
\expandafter\ifx\csname bibnamefont\endcsname\relax
  \def\bibnamefont#1{#1}\fi
\expandafter\ifx\csname bibfnamefont\endcsname\relax
  \def\bibfnamefont#1{#1}\fi
\expandafter\ifx\csname citenamefont\endcsname\relax
  \def\citenamefont#1{#1}\fi
\expandafter\ifx\csname url\endcsname\relax
  \def\url#1{\texttt{#1}}\fi
\expandafter\ifx\csname urlprefix\endcsname\relax\def\urlprefix{URL }\fi
\providecommand{\bibinfo}[2]{#2}
\providecommand{\eprint}[2][]{\url{#2}}

\bibitem[{\citenamefont{Suran and Zapesochnyi}(1975)}]{Suran1975}
\bibinfo{author}{\bibfnamefont{V.~V.} \bibnamefont{Suran}} \bibnamefont{and}
  \bibinfo{author}{\bibfnamefont{I.~P.} \bibnamefont{Zapesochnyi}},
  \bibinfo{journal}{Sov. Tech. Phys. Lett.} \textbf{\bibinfo{volume}{1}},
  \bibinfo{pages}{420} (\bibinfo{year}{1975}).

\bibitem[{\citenamefont{Suran and BondarÕ}(2009)}]{Suran2009}
\bibinfo{author}{\bibfnamefont{V.~V.} \bibnamefont{Suran}} \bibnamefont{and}
  \bibinfo{author}{\bibfnamefont{I.~I.} \bibnamefont{BondarÕ}},
  \bibinfo{journal}{Laser Physics} \textbf{\bibinfo{volume}{19}},
  \bibinfo{pages}{1502} (\bibinfo{year}{2009}).

\bibitem[{\citenamefont{Becker et~al.}(2012)\citenamefont{Becker, Liu, Ho, and
  Eberly}}]{Becker2012}
\bibinfo{author}{\bibfnamefont{W.}~\bibnamefont{Becker}},
  \bibinfo{author}{\bibfnamefont{X.}~\bibnamefont{Liu}},
  \bibinfo{author}{\bibfnamefont{P.~J.} \bibnamefont{Ho}}, \bibnamefont{and}
  \bibinfo{author}{\bibfnamefont{J.~H.} \bibnamefont{Eberly}},
  \bibinfo{journal}{Rev. Mod. Phys.} \textbf{\bibinfo{volume}{84}},
  \bibinfo{pages}{1011} (\bibinfo{year}{2012}).

\bibitem[{\citenamefont{Liontos et~al.}(2008)\citenamefont{Liontos, Cohen, and
  Bolovinos}}]{Liontos2008}
\bibinfo{author}{\bibfnamefont{I.}~\bibnamefont{Liontos}},
  \bibinfo{author}{\bibfnamefont{S.}~\bibnamefont{Cohen}}, \bibnamefont{and}
  \bibinfo{author}{\bibfnamefont{A.}~\bibnamefont{Bolovinos}},
  \bibinfo{journal}{J. Phys. B} \textbf{\bibinfo{volume}{41}},
  \bibinfo{pages}{045601} (\bibinfo{year}{2008}).

\bibitem[{\citenamefont{Motomura et~al.}(2009)\citenamefont{Motomura, Fukuzawa,
  Foucar, Liu, Pr\"{u}mper, Ueda, Saito, Iwayama, Nagaya, Murakami
  et~al.}}]{Motomura2009}
\bibinfo{author}{\bibfnamefont{K.}~\bibnamefont{Motomura}},
  \bibinfo{author}{\bibfnamefont{H.}~\bibnamefont{Fukuzawa}},
  \bibinfo{author}{\bibfnamefont{L.}~\bibnamefont{Foucar}},
  \bibinfo{author}{\bibfnamefont{X.-J.} \bibnamefont{Liu}},
  \bibinfo{author}{\bibfnamefont{G.}~\bibnamefont{Pr\"{u}mper}},
  \bibinfo{author}{\bibfnamefont{K.}~\bibnamefont{Ueda}},
  \bibinfo{author}{\bibfnamefont{N.}~\bibnamefont{Saito}},
  \bibinfo{author}{\bibfnamefont{H.}~\bibnamefont{Iwayama}},
  \bibinfo{author}{\bibfnamefont{K.}~\bibnamefont{Nagaya}},
  \bibinfo{author}{\bibfnamefont{H.}~\bibnamefont{Murakami}},
  \bibnamefont{et~al.}, \bibinfo{journal}{J. Phys. B}
  \textbf{\bibinfo{volume}{42}}, \bibinfo{pages}{221003}
  (\bibinfo{year}{2009}).

\bibitem[{\citenamefont{Liontos et~al.}(2010)\citenamefont{Liontos, Cohen, and
  Lyras}}]{Liontos2010}
\bibinfo{author}{\bibfnamefont{I.}~\bibnamefont{Liontos}},
  \bibinfo{author}{\bibfnamefont{S.}~\bibnamefont{Cohen}}, \bibnamefont{and}
  \bibinfo{author}{\bibfnamefont{A.}~\bibnamefont{Lyras}}, \bibinfo{journal}{J.
  Phys. B} \textbf{\bibinfo{volume}{43}}, \bibinfo{pages}{095602}
  (\bibinfo{year}{2010}).

\bibitem[{\citenamefont{Liu et~al.}(2010)\citenamefont{Liu, Ye, Liu, Rudenko,
  Tschuch, D\"{u}rr, Siegel, Morgner, Gong, Moshammer et~al.}}]{Liu2010}
\bibinfo{author}{\bibfnamefont{Y.}~\bibnamefont{Liu}},
  \bibinfo{author}{\bibfnamefont{D.}~\bibnamefont{Ye}},
  \bibinfo{author}{\bibfnamefont{J.}~\bibnamefont{Liu}},
  \bibinfo{author}{\bibfnamefont{A.}~\bibnamefont{Rudenko}},
  \bibinfo{author}{\bibfnamefont{S.}~\bibnamefont{Tschuch}},
  \bibinfo{author}{\bibfnamefont{M.}~\bibnamefont{D\"{u}rr}},
  \bibinfo{author}{\bibfnamefont{M.}~\bibnamefont{Siegel}},
  \bibinfo{author}{\bibfnamefont{U.}~\bibnamefont{Morgner}},
  \bibinfo{author}{\bibfnamefont{Q.}~\bibnamefont{Gong}},
  \bibinfo{author}{\bibfnamefont{R.}~\bibnamefont{Moshammer}},
  \bibnamefont{et~al.}, \bibinfo{journal}{Phys. Rev. Lett.}
  \textbf{\bibinfo{volume}{104}}, \bibinfo{pages}{173002}
  (\bibinfo{year}{2010}).

\bibitem[{\citenamefont{Delone and Krainov}(2000)}]{Delone2000}
\bibinfo{author}{\bibfnamefont{N.~B.} \bibnamefont{Delone}} \bibnamefont{and}
  \bibinfo{author}{\bibfnamefont{V.~P.} \bibnamefont{Krainov}},
  \emph{\bibinfo{title}{Multiphoton processes in atoms}}
  (\bibinfo{publisher}{Springer}, \bibinfo{year}{2000}).

\bibitem[{\citenamefont{Krausz and Ivanov}(2009)}]{Krausz2009}
\bibinfo{author}{\bibfnamefont{F.}~\bibnamefont{Krausz}} \bibnamefont{and}
  \bibinfo{author}{\bibfnamefont{M.}~\bibnamefont{Ivanov}},
  \bibinfo{journal}{Rev. Mod. Phys.} \textbf{\bibinfo{volume}{81}},
  \bibinfo{pages}{1163} (\bibinfo{year}{2009}).

\bibitem[{\citenamefont{Bernat et~al.}(1991)\citenamefont{Bernat, Bondar', and
  Suran}}]{Bernat1991}
\bibinfo{author}{\bibfnamefont{T.~T.} \bibnamefont{Bernat}},
  \bibinfo{author}{\bibfnamefont{I.~I.} \bibnamefont{Bondar'}},
  \bibnamefont{and} \bibinfo{author}{\bibfnamefont{V.~V.} \bibnamefont{Suran}},
  \bibinfo{journal}{Opt. Spectrosc.} \textbf{\bibinfo{volume}{71}},
  \bibinfo{pages}{22} (\bibinfo{year}{1991}).

\bibitem[{\citenamefont{Feldman et~al.}(1982)\citenamefont{Feldman, Krautwald,
  and Welge}}]{Feldman1982}
\bibinfo{author}{\bibfnamefont{D.}~\bibnamefont{Feldman}},
  \bibinfo{author}{\bibfnamefont{H.~J.} \bibnamefont{Krautwald}},
  \bibnamefont{and} \bibinfo{author}{\bibfnamefont{H.~J.} \bibnamefont{Welge}},
  \bibinfo{journal}{J. Phys. B} \textbf{\bibinfo{volume}{15}},
  \bibinfo{pages}{L529} (\bibinfo{year}{1982}).

\bibitem[{\citenamefont{Agostini and Petite}(1984)}]{Agostini1984}
\bibinfo{author}{\bibfnamefont{P.}~\bibnamefont{Agostini}} \bibnamefont{and}
  \bibinfo{author}{\bibfnamefont{G.}~\bibnamefont{Petite}},
  \bibinfo{journal}{J. Phys. B} \textbf{\bibinfo{volume}{17}},
  \bibinfo{pages}{L811} (\bibinfo{year}{1984}).

\bibitem[{\citenamefont{Agostini and Petite}(1985)}]{Agostini1985}
\bibinfo{author}{\bibfnamefont{P.}~\bibnamefont{Agostini}} \bibnamefont{and}
  \bibinfo{author}{\bibfnamefont{G.}~\bibnamefont{Petite}},
  \bibinfo{journal}{J. Phys. B} \textbf{\bibinfo{volume}{18}},
  \bibinfo{pages}{L281} (\bibinfo{year}{1985}).

\bibitem[{\citenamefont{Dexter et~al.}(1985)\citenamefont{Dexter, Jaffe, and
  Gallagher}}]{Dexter1985}
\bibinfo{author}{\bibfnamefont{J.~L.} \bibnamefont{Dexter}},
  \bibinfo{author}{\bibfnamefont{S.~M.} \bibnamefont{Jaffe}}, \bibnamefont{and}
  \bibinfo{author}{\bibfnamefont{T.~F.} \bibnamefont{Gallagher}},
  \bibinfo{journal}{J. Phys. B} \textbf{\bibinfo{volume}{18}},
  \bibinfo{pages}{L529} (\bibinfo{year}{1985}).

\bibitem[{\citenamefont{Petite and Agostini}(1986)}]{Petite1986}
\bibinfo{author}{\bibfnamefont{G.}~\bibnamefont{Petite}} \bibnamefont{and}
  \bibinfo{author}{\bibfnamefont{P.}~\bibnamefont{Agostini}},
  \bibinfo{journal}{J. Physique} \textbf{\bibinfo{volume}{47}},
  \bibinfo{pages}{795} (\bibinfo{year}{1986}).

\bibitem[{\citenamefont{Bondar' et~al.}(1988)\citenamefont{Bondar', Delone,
  Dudich, and Suran}}]{Bondar1988}
\bibinfo{author}{\bibfnamefont{I.~I.} \bibnamefont{Bondar'}},
  \bibinfo{author}{\bibfnamefont{N.~B.} \bibnamefont{Delone}},
  \bibinfo{author}{\bibfnamefont{M.~I.} \bibnamefont{Dudich}},
  \bibnamefont{and} \bibinfo{author}{\bibfnamefont{V.~V.} \bibnamefont{Suran}},
  \bibinfo{journal}{J. Phys. B} \textbf{\bibinfo{volume}{21}},
  \bibinfo{pages}{2763} (\bibinfo{year}{1988}).

\bibitem[{\citenamefont{Nakhate et~al.}(1991)\citenamefont{Nakhate, Ahmad,
  Razvi, and Saksena}}]{Nakhate1991}
\bibinfo{author}{\bibfnamefont{S.~G.} \bibnamefont{Nakhate}},
  \bibinfo{author}{\bibfnamefont{S.~A.} \bibnamefont{Ahmad}},
  \bibinfo{author}{\bibfnamefont{M.~A.~N.} \bibnamefont{Razvi}},
  \bibnamefont{and} \bibinfo{author}{\bibfnamefont{G.~D.}
  \bibnamefont{Saksena}}, \bibinfo{journal}{J. Phys. B}
  \textbf{\bibinfo{volume}{24}}, \bibinfo{pages}{4973} (\bibinfo{year}{1991}).

\bibitem[{\citenamefont{Tate et~al.}(1991)\citenamefont{Tate, Papaioannou, and
  Gallagher}}]{Tate1991}
\bibinfo{author}{\bibfnamefont{D.~A.} \bibnamefont{Tate}},
  \bibinfo{author}{\bibfnamefont{D.~G.} \bibnamefont{Papaioannou}},
  \bibnamefont{and} \bibinfo{author}{\bibfnamefont{T.~F.}
  \bibnamefont{Gallagher}}, \bibinfo{journal}{J. Phys. B}
  \textbf{\bibinfo{volume}{21}}, \bibinfo{pages}{1953} (\bibinfo{year}{1991}).

\bibitem[{\citenamefont{Haugen and Stapelfeldt}(1992)}]{Haugen1992}
\bibinfo{author}{\bibfnamefont{H.~K.} \bibnamefont{Haugen}} \bibnamefont{and}
  \bibinfo{author}{\bibfnamefont{H.}~\bibnamefont{Stapelfeldt}},
  \bibinfo{journal}{Phys. Rev. A} \textbf{\bibinfo{volume}{45}},
  \bibinfo{pages}{1847} (\bibinfo{year}{1992}).

\bibitem[{\citenamefont{Suran and BondarÕ}(2005)}]{Suran2005}
\bibinfo{author}{\bibfnamefont{V.~V.} \bibnamefont{Suran}} \bibnamefont{and}
  \bibinfo{author}{\bibfnamefont{I.~I.} \bibnamefont{BondarÕ}},
  \bibinfo{journal}{Optics and Spectroscopy} \textbf{\bibinfo{volume}{98}},
  \bibinfo{pages}{175} (\bibinfo{year}{2005}).

\bibitem[{\citenamefont{BondarÕ and Suran}(2003)}]{Bondar2003}
\bibinfo{author}{\bibfnamefont{I.~I.} \bibnamefont{BondarÕ}} \bibnamefont{and}
  \bibinfo{author}{\bibfnamefont{V.~V.} \bibnamefont{Suran}},
  \bibinfo{journal}{Optics and Spectroscopy} \textbf{\bibinfo{volume}{94}},
  \bibinfo{pages}{483} (\bibinfo{year}{2003}).

\bibitem[{\citenamefont{Ammosov et~al.}(1992)\citenamefont{Ammosov, Delone,
  Ivanov, Bondar, and Masalov}}]{Ammosov1992}
\bibinfo{author}{\bibfnamefont{M.~V.} \bibnamefont{Ammosov}},
  \bibinfo{author}{\bibfnamefont{N.~B.} \bibnamefont{Delone}},
  \bibinfo{author}{\bibfnamefont{M.~Y.} \bibnamefont{Ivanov}},
  \bibinfo{author}{\bibfnamefont{I.~I.} \bibnamefont{Bondar}},
  \bibnamefont{and} \bibinfo{author}{\bibfnamefont{A.~V.}
  \bibnamefont{Masalov}}, \bibinfo{journal}{Advances in Atomic, Molecular. And
  Optical Physics} \textbf{\bibinfo{volume}{29}}, \bibinfo{pages}{33}
  (\bibinfo{year}{1992}).

\bibitem[{\citenamefont{BondarÕ and Suran}(1998{\natexlab{a}})}]{Bondar1998}
\bibinfo{author}{\bibfnamefont{I.~I.} \bibnamefont{BondarÕ}} \bibnamefont{and}
  \bibinfo{author}{\bibfnamefont{V.~V.} \bibnamefont{Suran}},
  \bibinfo{journal}{JETP Letters} \textbf{\bibinfo{volume}{68}},
  \bibinfo{pages}{837} (\bibinfo{year}{1998}{\natexlab{a}}).

\bibitem[{\citenamefont{Bondar et~al.}(2000)\citenamefont{Bondar, Suran, and
  Dudich}}]{Bondar2000}
\bibinfo{author}{\bibfnamefont{I.~I.} \bibnamefont{Bondar}},
  \bibinfo{author}{\bibfnamefont{V.~V.} \bibnamefont{Suran}}, \bibnamefont{and}
  \bibinfo{author}{\bibfnamefont{M.~I.} \bibnamefont{Dudich}},
  \bibinfo{journal}{J. Phys. B} \textbf{\bibinfo{volume}{33}},
  \bibinfo{pages}{4243} (\bibinfo{year}{2000}).

\bibitem[{\citenamefont{BondarÕ and Suran}(1998{\natexlab{b}})}]{Bondar1998c}
\bibinfo{author}{\bibfnamefont{I.~I.} \bibnamefont{BondarÕ}} \bibnamefont{and}
  \bibinfo{author}{\bibfnamefont{V.~V.} \bibnamefont{Suran}},
  \bibinfo{journal}{Optics and Spectroscopy} \textbf{\bibinfo{volume}{85}},
  \bibinfo{pages}{327} (\bibinfo{year}{1998}{\natexlab{b}}).

\bibitem[{\citenamefont{Bondar et~al.}(2004)\citenamefont{Bondar, Suran, and
  Bondar}}]{Bondar2004}
\bibinfo{author}{\bibfnamefont{I.~I.} \bibnamefont{Bondar}},
  \bibinfo{author}{\bibfnamefont{V.~V.} \bibnamefont{Suran}}, \bibnamefont{and}
  \bibinfo{author}{\bibfnamefont{D.~I.} \bibnamefont{Bondar}},
  \bibinfo{journal}{Laser Physics} \textbf{\bibinfo{volume}{14}},
  \bibinfo{pages}{64} (\bibinfo{year}{2004}).

\bibitem[{\citenamefont{Kozlov}(1981)}]{Kozlov1981}
\bibinfo{author}{\bibfnamefont{M.~G.} \bibnamefont{Kozlov}},
  \emph{\bibinfo{title}{Absorption spectra of metallic vapors in vacuum
  ultraviolet}} (\bibinfo{publisher}{Nauka}, \bibinfo{address}{Moscow},
  \bibinfo{year}{1981}), \bibinfo{note}{see page 263 [in Russian]}.

\bibitem[{\citenamefont{Kotochigova and Tupizin}(1987)}]{Kotochigova1987}
\bibinfo{author}{\bibfnamefont{S.~A.} \bibnamefont{Kotochigova}}
  \bibnamefont{and} \bibinfo{author}{\bibfnamefont{I.~I.}
  \bibnamefont{Tupizin}}, \bibinfo{journal}{J. Phys. B}
  \textbf{\bibinfo{volume}{20}}, \bibinfo{pages}{4759} (\bibinfo{year}{1987}).

\bibitem[{\citenamefont{Bente and Hogervost}(1989)}]{Bente1989}
\bibinfo{author}{\bibfnamefont{E.~A. J.~M.} \bibnamefont{Bente}}
  \bibnamefont{and}
  \bibinfo{author}{\bibfnamefont{W.}~\bibnamefont{Hogervost}},
  \bibinfo{journal}{J. Phys. B} \textbf{\bibinfo{volume}{22}},
  \bibinfo{pages}{2679} (\bibinfo{year}{1989}).

\end{thebibliography}
\end{document}